\documentclass[%
 reprint,%
 amssymb, amsmath,%
 aip,cha,%
]{revtex4-1}
\usepackage{graphicx}
\usepackage{docs}%
\usepackage{bm}%
\usepackage[colorlinks=true,linkcolor=blue]{hyperref}%
\expandafter\ifx\csname package@font\endcsname\relax\else
 \expandafter\expandafter
 \expandafter\usepackage
 \expandafter\expandafter
 \expandafter{\csname package@font\endcsname}%
\fi
\begin{document}

\title{Quantum magnetoresistance of the PrFeAsO oxypnictide}%
\author{D. Bhoi}
\email{dilipkumar.bhoi@saha.ac.in}
\author{P. Mandal}%
 \email{prabhat.mandal@saha.ac.in}
\affiliation{Saha Institute of Nuclear Physics, 1/AF Bidhannagar,Calcutta 700 064, India
}%
\author{P. Choudhury}

\affiliation{Central Glass and Ceramic Research Institute, 196 Raja S. C. Mullick Road, Calcutta  700 032, India
}%
\author{S. Pandya}
\author{V.Ganesan}
\affiliation{UGC-DAE Consortium for Scientific Research, University Campus, Khandwa Road, Indore 452
017, India}

\date{\today}%
\begin{abstract}
We report the observation of an unusual $B$ dependence of transverse magnetoresistance (MR) in the PrFeAsO, one of the parent compound of pnictide superconductors. Below the spin density wave transition, MR is large, positive and increases with decreasing temperature. At low temperatures, MR increases linearly with $B$ up to 14 T. For $T$$\geq$40 K, MR vs $B$ curve develops a weak curvature in the low-field region which indicates a crossover from $B$ linear to $B^2$ dependence as $B$$\rightarrow$0. The $B$ linear MR originates from the Dirac cone states and has been explained by the quantum mechanical model proposed by Abrikosov.
\end{abstract}

\maketitle
The Fe-pnictide high-temperature superconductors have attracted much interest because of the interplay of multi-band structure of Fermi surface and antiferromagnetism mediated by the magnetic Fe ions \cite{hosno}. The undoped compounds $R$FeAsO ($R$$=$La, Ce, Pr, Nd and Sm) and $A$Fe$_2$As$_2$ ($A$$=$Ba, Ca, Sr, Eu) are semimetal and exhibit a structural phase transition at $T_S$ followed by an antiferromagnetic spin-density wave (SDW) magnetic ordering at $T_N$ \cite{john}. This SDW has been argued to arise from Fermi surface nesting of the hole and electron Fermi surface sheets, and/or from one-electron band effects \cite{johan}. The theoretical calculation shows that the combination of the physical symmetry and topology of the band structure naturally stabilizes a gapless nodal SDW state with Dirac nodes for the parent compounds of pnictide \cite{ran,harrison}. Angle-resolved photoemission spectroscopy  measurement on BaFe$_2$As$_2$ confirms a Dirac-type dispersion in the antiferromagnetic phase \cite{richard}.

Dirac type dispersion was first observed in graphene, where the energy spectrum is linear in momentum at the corners of the first Brillouin zone, so that the electrons at low energies can be described by the Dirac equation \cite{nose,zhang}. Such a unique energy band structure gives rise to several characteristic transport properties \cite{geim}. In this letter, we report the investigation on the magnetotransport properties of PrFeAsO sample, one of the parent compound of 1111 oxypnictides. We observe a large positive magnetoresistance (MR) below the SDW transition temperature. At low temperature, the MR follows a linear field dependence up to 14 T magnetic field. For $T$$\geq$40 K, MR vs $B$ curve develops a weak curvature in the low-field region which indicates a crossover from $B$ linear to $B^2$ dependence. The $B$ linear dependence of MR is analyzed by the quantum mechanical model developed by Abrikosov \cite{abri1,abri2}.

Polycrystalline PrFeAsO sample has been prepared by standard solid state reaction method as described in our earlier reports \cite{bhoi}. The phase purity of the sample was checked by powder X-ray diffraction method and no trace of the impurity phase has been detected. The scanning electron microscope (SEM) (SUPRA, 35 VP, Carl Zeiss) image reveals well connected platelet crystallites. The energy dispersive X-ray spectra obtained from grains of different size and morphology reveal that the stoichiometric ratio Pr:Fe:As:O is close to the nominal composition PrFeAsO. The electrical resistivity measurement was carried out in 14 T physical property measurement system (Quantum Design). The Hall coefficient ($R_H$) was measured at some selected temperatures by sweeping the field from -7 T to 7 T in a superconducting magnet (Oxford Instruments).

\begin{figure}[b]
  \includegraphics[width=0.45\textwidth]{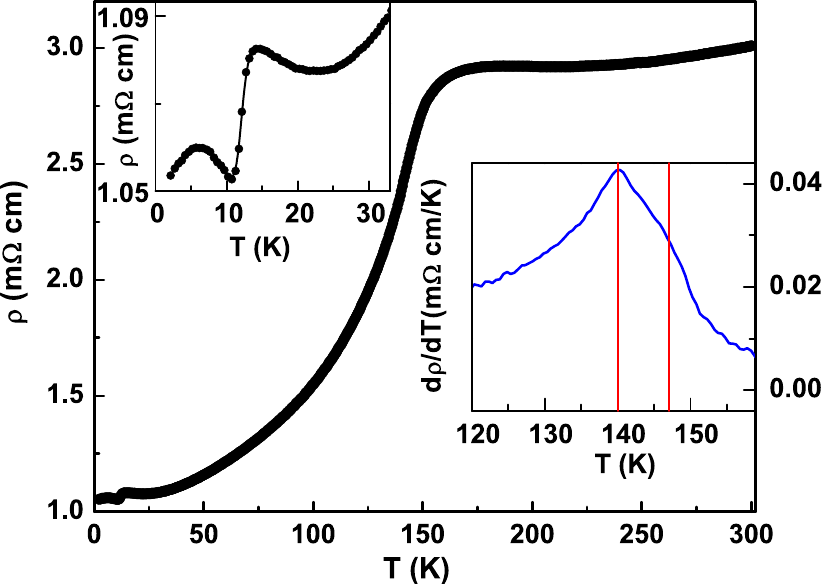}\\
  \caption{(Color online) Temperature dependence of $\rho$ for the PrFeAsO sample. Lower inset: $d\rho/dT$ versus $T$ curve of the sample displaying the structural and magnetic transitions around 150 K. Upper inset: low-temperature behavior of $\rho$ displaying the resistivity minimum at around 23 K and AFM ordering of the Pr moments at $T_N^{\rm{Pr}}$.}\label{Fig.1}
\end{figure}

Figure 1 shows the resistivity of the undoped PrFeAsO sample. Upon cooling, the resistivity decreases slowly down to 214 K, indicating metallic behavior. The broad peak at around 150 K is associated with both the structural phase transition from tetragonal to orthorhombic symmetry and the SDW magnetic transition. Two overlapping peaks are identified in the d$\rho$/d$T$ vs $T$ plot (lower inset of Fig.1), one due to the crystallographic distortion at $T_S$ (147 K) and the other due to the onset of long-range SDW magnetic ordering of Fe moments at $T_{N}$ (140 K). Below $T_{N}$, the resistivity of the sample initially decreases with the decrease of $T$ and then passes through a shallow minimum at $\sim$23 K (upper inset of Fig. 1). As $T$ is lowered further, $\rho$ exhibits a weak upturn before a sharp drop of 4\% occurs due to the AFM ordering of 4$f$ electrons of the Pr ion. The transition temperature $T_{N}^{\rm{Pr}}$ determined from the peak in the d$\rho$/d$T$ vs $T$ curve is $\sim$12 K, which is close to that reported from neutron diffraction, magnetic susceptibility and other measurements \cite{kimber,rotundu}. One can see that $\rho$($T$) curve exhibits another weak maximum at around 5 K. This peak in $\rho$ is due to the Fe-spin reorientation induced by the Pr-AFM sublattice \cite{macg}.

Figure 2 shows some representative plots of $B$ dependence of the transverse magnetoresistance (MR) at different temperatures, where MR($B$)$=$$\Delta\rho/\rho(0)$=$\left[\rho(B)-\rho(0)\right]/\rho(0)$. Above 150 K, the value of MR is very small ($<$1\%) even at a field of 14 T. Below 150 K, the MR increases rapidly with decreasing temperature and it reaches 63\% at 5 K and 14 T. It is interesting to note that MR increases linearly with $B$ in the entire field range below 40 K. The MR at 5 K increases linearly with $B$ both below and above 6.4 T but with slight different slopes [Fig. 2(b)]. This behavior is signalling a weak metamagnetic transition similar to that observed in Ce-based pnictide \cite{jesche}. For $T$$\geq$40 K, $\Delta\rho/\rho(0)$ vs $B$ curve develops a weak curvature in the low-field region which indicates a crossover to a quadratic behavior as $B$$\rightarrow$0. The $B^2$ dependence of MR becomes more evident on examining the $d$MR($B$)/$dB$ at 40 K as shown in the Fig. 2(c), where $d$MR($B$)/$dB$ saturates in the high field region but it shows an approximate linear $B$ dependence below a critical field $B_c$$\sim$1.6 T.  A crossover from linear $B$ dependence at high field to superlinear $B$ dependence at low field of MR has been reported for several materials \cite{hu,adam}.

\begin{figure}[t]
  \includegraphics[width=0.45\textwidth]{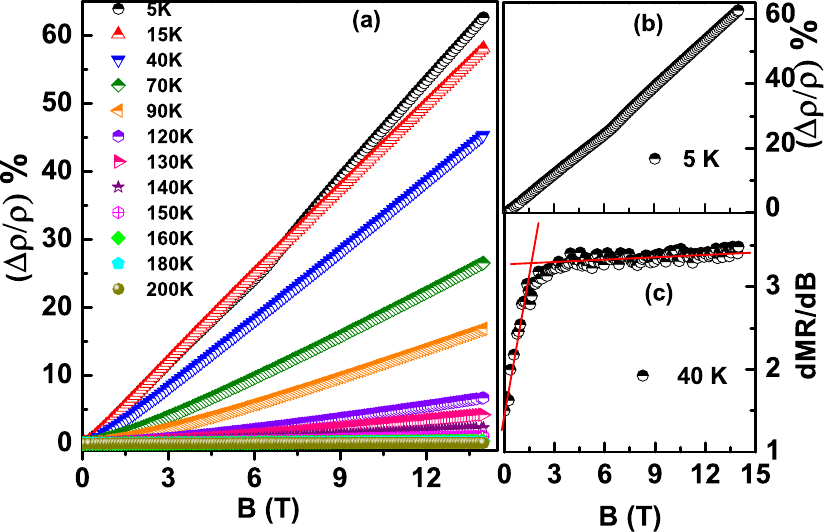}\\
  \caption{(Color online) (a) Field dependence of the MR [=$\Delta\rho/ \rho(0)$] for the PrFeAsO sample at different temperatures. (b) MR at 5 K increases linearly with $B$ up to 6.4 T and it follows a steeper slope above this field. (c) $d$MR/$dB$ at 40 K saturates above $B \ge$ 1.6 T exhibiting the linear $B$ dependence in high magnetic field.}\label{Fig.2}
\end{figure}

According to semiclassical theory, the magnetoresistance in a metal with nonequal densities of electrons and holes behaves as
\[\Delta\rho\sim \left\{\begin{array}{ll}\rho(0)(\omega_c\tau)^2 & {\omega_c\tau\ll1}\\
                                         \rho(0) & {\omega_c\tau\gg1},\end{array}\right.\]

where, $\omega_c$=$\frac{eB}{m_1}$ is the cyclotron frequency, $m_1$ is the cyclotron mass, $\tau$ is the collision time, and $\rho(0)$ is the resistivity at zero field. This means that first it grows quadratically with field and then reaches saturation. In multi-band systems also, $\Delta\rho/\rho(0)$ varies quadratically with $B$ in the low-field region. Hence, the large value and the linear $B$ dependence of MR in the present compound have some different origin. There exists a certain possibility of a linear magnetoresistance, namely, in a polycrystalline metal with an open Fermi surface but this is definitely not the  case for the present sample. A number of semi-metals and narrow-gap semiconductors also show large positive and non-saturating linear magnetoresistance over a wide range of $B$ and $T$ \cite{xu,yang,hu}. A quantum description for this magnetotransport phenomenon was developed by Abrikosov \cite{abri1,abri2}.

Quantum effects become noticeable when the Landau levels (LL) associated with the electron orbits are distinct: $\hbar\omega_c$$\gg$$k_BT$. At very low temperature and high fields, it is possible to reach the `extreme quantum limit', where $\hbar \omega_c$ can exceed the Fermi energy $E_F$ and electrons can coalesce into the lowest LL of the transverse motion in $B$. In this limit, Abrikosov calculated the magnetoresistivity \cite{abri1,abri2}, considering $B$ is along $z$-axis, as
\begin{equation}
\rho_{xx} = \rho_{yy} = \frac{N_iB}{\pi n_0^2e},
\end{equation}
where $\rho_{xx}$ and $\rho_{yy}$ are transverse components of the resistivity tensor, $n_0$ the electron density, and $N_i$ the concentration of scattering centers. Equation (1) is valid under the condition, $n_0$$\ll$$\left(\frac{eB}{\hbar}\right)^{3/2}$. This condition is satisfied in semimetallic materials like Bi \cite{yang} and narrow gap semiconductor like Ag$_{2+\delta}$Se and Ag$_{2+\delta}$Te \cite{xu}, and Eqn. (1) successfully explains the observed $B$ linear dependence of MR. In Fig. 3, we have plotted the temperature dependence of 1/$|eR_H|$, where $R_H$ is the Hall coefficient at 7 T. Though pnictides are multiband system, assuming a simple one band model we can evaluate the upper limit for carrier concentration as 1/$|eR_H|$. With decreasing $T$, the carrier concentration decreases and at the SDW transition it falls by a factor of seven. The carrier concentration at 5 K is roughly three orders of magnitude smaller than the room temperature value. Thus, it appears that for the present sample or pnictides the carrier density is high to satisfy the condition $n_0$$\ll$$\left(\frac{eB}{\hbar}\right)^{3/2}$. However, it is possible that there exists a small pocket on the Fermi surface where the quantum condition is satisfied. Indeed, this has been observed for several materials like InSb \cite{hu}, $R$Sb$_2$ ($R$ = La-Nd, Sm) \cite{budko} and multilayer epitaxial graphene \cite{adam}. In this pocket of Fermi surface, the relationship between energy and momentum is linear leading to zero effective mass of electron \cite{abri2}.
\begin{figure}
  \includegraphics[width=0.45\textwidth]{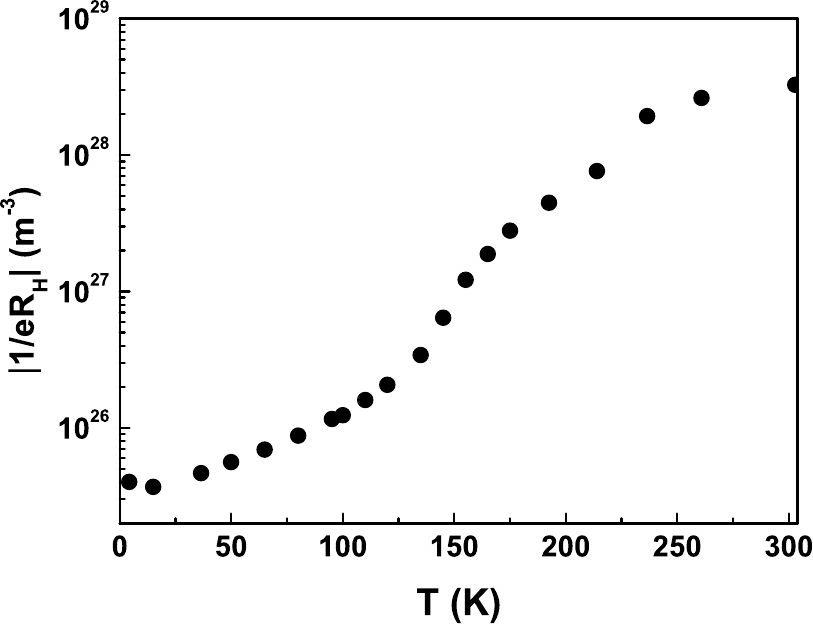}\\
  \caption{(Color online) Temperature dependence of the 1/$|eR_H|$ at 7 T for the PrFeAsO sample.}\label{Fig.3}
\end{figure}

In the parent compound of pnictides, a Dirac surface state is created due to the SDW band reconstruction and the apex of the Dirac dispersion intersects the Fermi energy, giving rise to electron and hole pockets \cite{ran,harrison,richard}. In the Dirac state, the quantum condition is satisfied for InSb, $R$Sb$_2$ and multilayer epitaxial graphene \cite{hu,adam,budko}. In a parabolic band, the LL splitting is proportional to the first order of $B$; i.e., $\Delta_n$$=$$\hbar eB/m^\ast$, where $m^\ast$ is the effective mass defined by the curvature of the band. In contrast, the LL splitting in a Dirac state scales with the square root of $B$; i.e., $\Delta_n$$=$$\pm v_F \sqrt{2\hbar eB|n|}$, where $v_F$ is the Fermi velocity. This implies that the energy scale associated with the electron in the Dirac state is quite different from the electron in a parabolic band. For example, the energy scale for the Dirac fermions with $v_F$$\sim10^6$ m/s is more than 2 orders of magnitude larger than the cyclotron energy for the electrons in the parabolic band at $B$$=$1 T. This makes quantum phenomenon observable even at high temperatures and in low or moderate field strength for the Dirac fermions. The linear MR for the entire $B$ range observed in low temperature, therefore, can originate from the Dirac cone states. For $T$$\geq$40 K, there may have two possibilities for the occurrence of crossover from linear to quadratic $B$ dependence in MR with decreasing $B$. One is the increase of $n_0$ with increasing $T$ as observed in Fig. 3. The strong $T$ dependence of 1/$|eR_H|$ indicates the increase of the carrier concentration in the Dirac states. Consequently, the condition $n_0$$\ll$$\left(\frac{eB}{\hbar}\right)^{3/2}$ may not be valid at high temperatures and in the low-field limit. Another possibility is the involvement of two or three Landau levels due to the thermal and collision broadening with increasing $T$ similar to that observed in InSb \cite{hu}.

In summary, we have investigated the field and temperature dependence of MR in PrFeAsO oxypnictide. Below the SDW transition, MR increases with decreasing temperature and reaches 63\% at 5 K and 14 T. MR shows a linear $B$ dependence up to 14 T in the low temperature region due to the carriers in the Dirac cone in the SDW state. For $T \geq$ 40 K, a crossover from a linear $B$ dependence to a quadratic $B$ dependence in MR appears with decreasing $B$.

The authors would like to thank A. Pal, S. Banerjee, A. Midya and N. Khan for technical help. V. Ganesan would also like to thank DST, India for financial assistance for the 14 T physical property measurement system facility at CSR, Indore.


\begin{thebibliography}{99}

\bibitem{hosno} K. Ishida, Y. Nakai, and H. Hosono, J. Phys. Soc. Jpn. \textbf{78}, 062001 (2009).

\bibitem{john} David C. Johnston, Advances in Physics \textbf{ 59}, 803 (2010).

\bibitem{johan} M. D. Johannes and I. I. Mazin, Phys. Rev. B \textbf{79}, 220510(R) (2009).

\bibitem{ran} Y. Ran, F. Wang, H. Zhai, A. Vishwanath, and D.-H. Lee, Phys. Rev. B \textbf{79}, 014505 (2009).

\bibitem{harrison} N. Harrison and S. E. Sebastian, Phys. Rev. B \textbf{80}, 224512 (2009).

\bibitem{richard} P. Richard, K. Nakayama, T. Sato, M. Neupane, Y.-M. Xu, J. H. Bowen, G. F. Chen, J. L. Luo, N. L. Wang, X. Dai, Z. Fang, H. Ding,
and T. Takahashi, Phys. Rev. Lett. \textbf{104}, 137001 (2010).

\bibitem{nose} K. S. Novoselov, A. K. Geim, S. V. Morozov, D. Jiang, M. I. Katsnelson, I. V. Grigorieva, S. V. Dubonos, and A. A. Firsov, Nature (London) \textbf{438}, 197 (2005).

\bibitem{zhang} Yuanbo Zhang, Yan-Wen Tan, Horst L. Stormer, and Philip Kim, Nature (London) \textbf{438}, 201 (2005).

\bibitem{geim} A. K. Geim and K. S. Novoselov, Nature Mater. \textbf{6}, 183 (2007).

\bibitem{abri1}	A. A. Abrikosov, Sov. Phys.(JETP) \textbf{29}, 746 (1969).

\bibitem{abri2}	A. A. Abrikosov, Phys. Rev. B \textbf{58}, 2788 (1998); A. A. Abrikosov, Europhys. Lett. \textbf{49}, 789 (2000).

\bibitem{bhoi} D. Bhoi, P. Mandal, and P. Choudhury, Supercond. Sci. Technol. \textbf{21} 125021 (2008);D Bhoi, P. Mandal and P. Choudhury, Physica C \textbf{468}, 2275 (2008).

\bibitem{kimber} S. A. J. Kimber, D. N. Argyriou, F. Yokaichiya, K. Habicht, S. Gerischer, T. Hansen, T. Chatterji, R. Klingeler,C. Hess, G. Behr, A. Kondrat, and B. B\"{u}chner, Phys. Rev. B \textbf{78}, 140503(R)(2008).

\bibitem{rotundu} C. R. Rotundu, D. T. Keane, B. Freelon, S. D. Wilson, A. Kim, P. N. Valdivia, E. Bourret-Courchesne, and R. J. Birgeneau,
Phys. Rev. B \textbf{80}, 144517 (2009).

\bibitem{macg} M. A. McGuire, R. P. Hermann, A. S. Sefat, B. C. Sales, R. Jin, D. Mandrus, F. Grandjean and G. J. Long,
New Journal of Physics \textbf{11}, 025011 (2009).

\bibitem{jesche} A. Jesche, C. Krellner, M. de Souza, M. Lang and C. Geibel, New Journal of Physics \textbf{11}, 103050 (2009).

\bibitem{hu} J. S. Hu and T. F. Rosenbaum, Nat. Mater. \textbf{7}, 697 (2008).

\bibitem{adam} A. L. Friedman, J. L. Tedesco, P. M. Campbell, J. C. Culbertson, E. Aifer, F. K. Perkins, R. L. Myers-Ward, J. K. Hite, C. R. Eddy,
Jr., G. G. Jernigan, and D. K. Gaskill, Nano Lett. \textbf{10}, 3962 (2010).

\bibitem{xu} R. Xu, A. Husmann, T. F. Rosenbaum, M.-L. Saboungi, E. J. Enderby and P. B. Littlewood, Nature \textbf{390}, 57 (1997).

\bibitem{yang} F. Y. Yang, K. Liu, K. Hong, D. H. Reich, P. C. Searson and C. L. Chien, Science \textbf{284}, 1335 (1999).

\bibitem{budko} S. L. Bud'ko, P. C. Canfield, C. H. Mielke, and A. H. Lacerda, Phys. Rev. B \textbf{57}, 13624 (1998).

\end{thebibliography}
\end{document}